# AB Percolation in Semiconductor Alloys

Rohit Garg* and Kailash C. Rustagi[#]
Physics Department, Indian Institute of Technology Bombay, Mumbai 400076, India

ABSTRACT

We show that, at the percolation threshold, *two* percolating AB clusters form in the diamond structure simultaneously instead of the usual one in the common site percolation problems. The structure of these two clusters and the related physical properties of semiconductor mixed crystals are discussed.

PACS: 64.60.ah;61.43.Dq;61.66.Dk

___________________________________________________________________

*electronic address: rohit_garg@phys.iitb.ac.in
#electronic address: rustagi@iitb.ac.in

The percolation phenomena are believed to impact many physical properties of disordered solids [1-4]. In tetrahedrally bonded mixed crystals of the type $A_xB_{1-x}C$ (III-V or II-VI compounds) percolation related phenomena have been invoked to explain the observed departures from the predictions of the virtual crystal approximation. For mixed crystals of the type $A_xB_{1-x}C$ the atoms A and B occupy sites on a face centered cubic (FCC) lattice of the zinc blende structure while the other FCC lattice is always occupied by the C atoms. For $x \geq x_f$, the site percolation threshold for the FCC lattice, wall-to-wall connected networks or percolating clusters of AC bonds occur in the limit of infinite sample size while those of the AB bonds occur for all $x \leq (1-x_f)$. Bellaiche et al found an anomaly in the concentration dependence of the lattice constant in GaAsN[5]. Kim *et al* [6] invoked percolating low potential channels to interpret the anomalous tunneling of electrons through thick AlGaAs barriers. In a series of papers Pages *etal* [7] have interpreted the zone centre phonon spectra of mixed compound semiconductor crystals. In the much simpler system of IV-IV mixed semiconductor crystals like Si-Ge the percolating clusters have a larger variety. While Si-Si (or Ge-Ge) percolating clusters exist in $Si_x Ge_{(1-x)}$ crystals for $x \geq x_d$ (or $x \leq (1-x_d)$), where $x_d$ is the percolation threshold for the diamond structure; percolating clusters of Si-Ge can exist in a fairly large range of concentration $x_c \leq x \leq (1-x_c)$. It is expected that these percolating clusters will have a significant effect on the structural, vibrational and optical properties of these mixed crystals. We present a calculation of the percolation threshold $x_c$ and investigate the structure of these percolating clusters. We also find that in the percolation range $x_c \leq x \leq (1-x_c)$, two unconnected AB percolating clusters always exist. In contrast, there is always a unique percolating cluster in the standard lattice percolation problem. Physical implications of this result are then discussed.

The occurrence of percolating clusters in which successive sites are alternately occupied by A and B atoms, respectively, was called AB percolation by Halley[8] who also estimated the threshold for several simple lattices. Although, the claim that this kind of percolation does not occur on two-dimensional square lattice attracted some attention [9] there have been only a few investigations of AB percolation. For example, Silverman and Adler[10] have reported the threshold for finding percolating clusters of GaAs in mixed crystals of $(GaAs)_xGe_{1-2x}$ obtained by substituting a fraction x of Ge-Ge bonds by GaAs bonds. This is an approximation for the situation in which Ga and As are randomly substituted on Ge sites since in the latter case Ga-Ga and As-As bonds could also occur[11]. More generally, Ioselevich[12] found an approximation for the percolation threshold for polychromatic percolation in the context of random site hopping transport in polymers. Similarly, Harreis and Bauer [13] have reported percolation threshold for site bond percolation on a simple cubic lattice. Although it is obvious that several variants of polychromatic percolation exist, in this report we restrict ourselves to looking for AB percolating clusters where A and B are randomly substituted on a diamond structure.

We first calculated the number of percolating AB clusters as a function of *x*, the fraction of sites occupied by B atoms in a diamond structure sample with N×N×N cubic unit cells. The algorithm used to search for a percolating cluster was the breadth-first- search algorithm. It was tested for accuracy on several known two and three dimensional lattices. A percolating cluster was taken as one which connects opposite walls in all 3 directions. The average number of percolating clusters, averaged over 100 samples each, is shown in Fig 1 as a function of x, for N=98,198,298,398 and 498, respectively. The calculated points are well approximated by a *tanh* form, from which we extracted $x_h(N)$, the value of x for which the average number of percolating clusters is 1 for the NxNxN sample. The main result here is that in the percolating regime the number of percolating clusters is always 2 instead of the normal 1 in the standard percolation problem. To understand that this difference between AB percolation and the standard percolation problem, let us call the two FCC sub-lattices of the diamond structure as white and black sub-lattices, respectively. The nearest neighbors of every atom on white sub-lattice lie on the black sub-lattice and similarly the nearest neighbors of every atom on black sub-lattice lie on the white sub-lattice. One of the two percolating clusters has A on white and B on black sub-lattice while the other has B on white and A on black sub-lattice. Obviously the two clusters cannot share any atom in common and are therefore disconnected. We also observe that the two clusters are fractal structures obtained by removing a fraction of atoms from two *distinct* AB crystals which are inversion images of each other. The existence of two percolating clusters is thus a necessary consequence of the fact that the diamond structure has an inversion symmetric structure which can be obtained from two distinct zinc blende structures in the limit of their two atoms becoming identical.

To obtain a reliable estimate of the percolation threshold we use the usual finite length scaling procedure [1,10]. The difference in threshold $x_h - x_c$ as well as the width $\gamma(N)$ scale as $N^{-(1/\nu)}$ where $\nu$, the universal exponent is estimated to be 0.89 for 3 dimensional lattices[14]. In Fig 2 $x_h(N)$ is plotted as a function of $N^{-(1/\nu)}$, where a good fit is obtained for $x_h(N) = x_c + AN^{-(1/\nu)}$ with $x_c = 0.25637$. A power law fit with variable exponent gave the exponent as 1.02 instead of 0.89 without a significant improvement in the quality of the fit. Similarly, the width $\gamma(N)$ is also found to obey the expected scaling form, as shown in Fig 3 We note that our value of the percolation threshold $x_c$=0.25637 is only slightly higher than the estimate given by Ioselevich [12] $x_d = \sqrt{x_c(1-x_c)}$.

Since two percolating clusters appear at the percolation threshold it is interesting to examine the relative number of atoms in the two clusters. The difference between the average number of atoms in the two clusters is related to the fluctuations in the cluster size, which diverges at the percolation threshold for infinite lattice. We show this behavior for the largest sample size N=498 in Fig 4, where the standard deviation in the number of atoms is plotted as function of fraction *x*. For *x*>*x_h* this diverges ~(*x*-*x_h*)$^{-\eta}$, where $\eta = 0.75 \pm 0.11$.

In view of the apparent similarity between this problem and that considered by Silverman and Adler[10], it is interesting to compare the diatomic substitution case with ours. First, we note that even in that case two clusters would be found in the percolation regime if Ga and As were substituted randomly on Ge lattice even with the provision that Ga-Ga and As-As bonds were forbidden. This follows, because when we substitute a Ge-Ge bond by a GaAs bond there are two possible ways in which we can do that corresponding to Ga on the white sub-lattice or Ga on the black sub-lattice. However, if we choose to substitute Ge-Ge bonds by GaAs bonds in such a way that all Ga atoms always lie on the same FCC sub-lattice, only one kind of GaAs clusters can form. But in that case, percolating GaAs clusters exist for all *x*> *x_c*, the percolation threshold which occupies the full sample in the limit of *x*=1. In contrast, if we substitute Ga and As randomly for Ge atoms, the fully substituted sample containing 50% atoms of Ga and As will have four percolating clusters- one each of Ga and As and two with GaAs connected bonds. The percolation thresholds in the two cases are, however, similar.

Another related problem is that of percolating clusters in quaternary mixed semiconductors like $Ga_{1-x}Al_xAs_{1-y}P_y$ where we may assume that Ga and Al are randomly distributed on one FCC sub lattice and As and P on the other. One can then look for thresholds for finding percolating clusters of GaAs, GaP, AlAs and AlP. For x=y, GaAs percolating clusters will exist for $x \leq (1-x_c)$ while AlP percolating clusters of AlP will exist for $x \geq x_c$.

Finally, we look at possible effects of AB percolation on the physical properties of Si-Ge mixed crystals. For $Si_xGe_{(1-x)}$ as many as 4 distinct percolating clusters exist in the range $x_d \leq x \leq (1-x_d)$. Two of them are Si-Ge clusters of roughly equal average size while the other two are Si-Si and Ge-Ge percolating clusters. While the two Si-Ge percolating clusters do not share a common atom both Si-Si and Ge-Ge clusters may share common atoms with the Si-Ge clusters. Since every atom on the boundary of a Ge-Ge cluster has at least one Si nearest neighbor one expects that a Ge-Ge cluster would contact with a Si-Ge percolating cluster at several points. At $x$=0.5 the probability of forming the Si-Si bond is the same as that for forming a Si-Ge bond. But since Si-Ge bonds are shared between two Si-Ge percolating clusters, we expect that each of the two Si-Ge clusters is about half the size of Si-Si or Ge-Ge clusters.

Now, it is well known that Si-Si, Si-Ge and Ge-Ge bond lengths are nearly preserved in the mixed crystals of SiGe [15]. Since the structure of percolating clusters includes blobs, nodes and links at several length scales [13,16], it is indicated that there exist meso-scale regions over each of which the bond length remains nearly constant. Also, the connectivity of similar bonds plays an important role in determining the vibrational spectra [7]. Si-Ge percolating clusters are therefore expected to play an important role in interpreting the rich vibrational spectra of Si-Ge mixed crystals.

To conclude, we have shown that for AB percolation on diamond structure there are two independent percolating clusters in the percolation range $x_c \leq x \leq (1-x_c)$. Some physical implications of this in the context of SiGe mixed crystals are also described.

We are indebted to Prof Deepak Dhar for important suggestions and a critical reading of the manuscript. It is a pleasure to thank Prof Olivier Pages for stimulating discussions.

References:
1. D.Stauffer and A.Aharony , *Introduction to Percolation Theory*, Taylor and Francis, London(1994), D. Stauffer in *Quantum and Semi-classical Percolation and Breakdown in Disordered Solids*(Lecture notes in Physics vol **762**) edited by A. K. Sen, K. K. Bardhan, B. K. Chakrabarti Springer Berlin(2009)
2. H. E. Stanley, J. S. Andrade Jr., S. Havlinc, H.A. Makse, and B. Sukie,Physica A**266**,5(1999)
3. B.I.Shklovskii and A.L.Efros *Electronic Properties of Doped Semiconductors* , Springer, Berlin (1984)
4. J. Osorio-Guillén, S. Lany, S. V. Barabash, and A. Zunger, Phys. Rev. B **75**,184421(2007)
5. L. Bellaiche, S.-H. Wei, and Alex Zunger, Phys. Rev. B **54**, 17568 (1996)
6. D.S. Kim, H.S. Ko, Y.M. Kim, S.J. Rhee, S.C. Hohng, Y.H. Yee, W.S. Kim, J.C. Woo, H.J. Choi, J. Ihm, D.H. Woo and K.N. Kang, Phys RevB **54**,14580(1996), P. H. Song and D. S. Kim Phys. Rev. B **54**, R2288 (1996)
7. O. Pagès, A. V. Postnikov, M. Kassem, A. Chafi, A. Nassour, and S. Doyen,Phys. Rev. B **77**, 125208 (2008) , and references therein.
8. JW Halley, in *Percolation Structures and Processes*, G.Deutscher,R.Zallen,and J.Adler eds, Hilger Bristol(1983)
9. T.Luczak and J.C. Wierman J.Phys. A:Math Gen **22**,185(1989)
10. A.Silverman and J.Adler Phys. Rev. B **42**,1369(1990)
11. R.Oso'rio and S.Froyen, Phys. Rev. B **47**,1889(1993), H.Holloway, Phys.Rev. B **66**,075313(2002)
12. A.S.Ioselevich, Phys Rev Lett **74**,1411(1995),
13. H.M. Harreis and W. Bauer,  Phys Rev B **62**,8719(2000)
14. C.-Y.Lin and C.-K.Hu, Phys.Rev. E **58**,1521(1998)
15. M.M. Rieger and P. Vogl, Phys. Rev. B **48**, 14 276 (1993)
16. H.J.Herrmann and H.E.Stanley Phys. Rev. Lett. **53**,1121(1984)

Figures with captions:

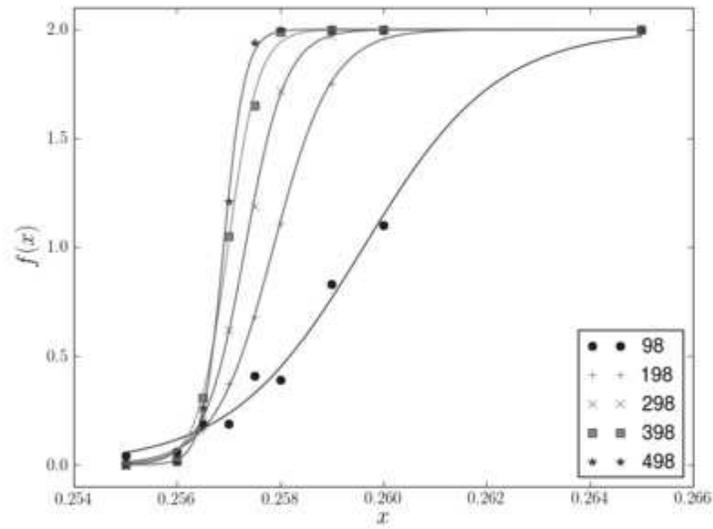

Fig 1: Average number of percolating clusters as function of fractional concentration x for sample size N=98(dots), 198(pluses), 298(crosses), 398(squares), and 498(asterisks'), respectively

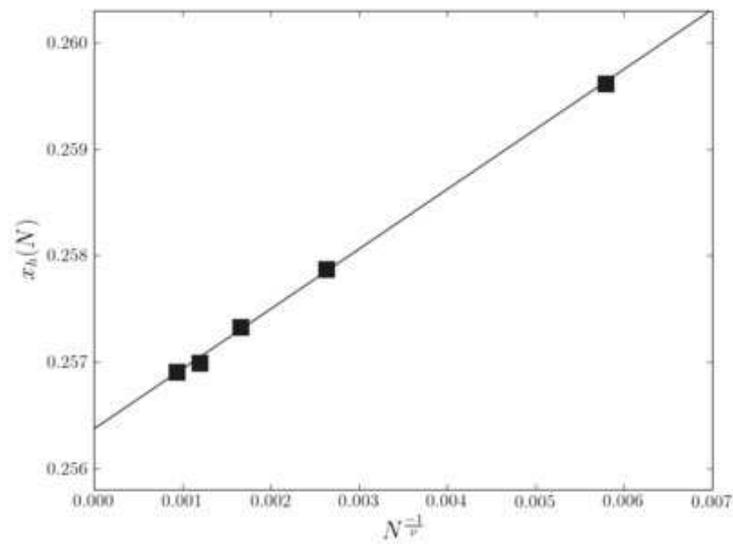

Fig 2: The threshold concentration $x_h$ as a function of $N^{-(1/\nu)}$.

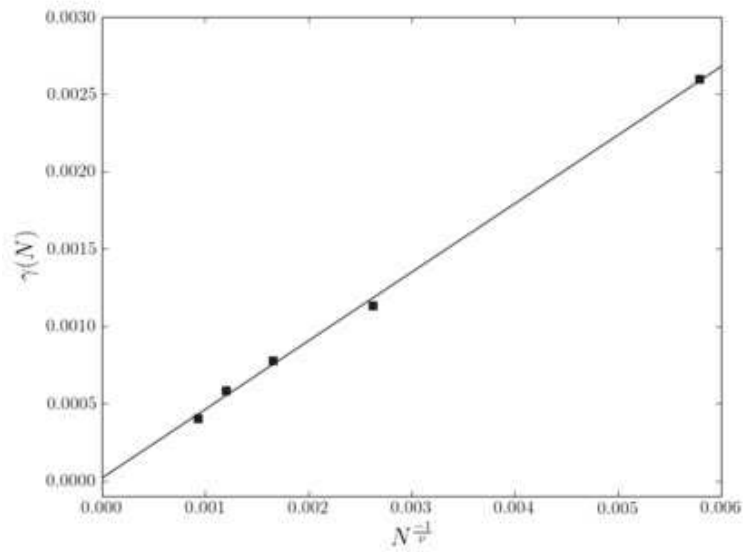

Fig3: The width $\gamma(N)$ as a function of $N^{-(1/\nu)}$ with $\nu=0.89$.

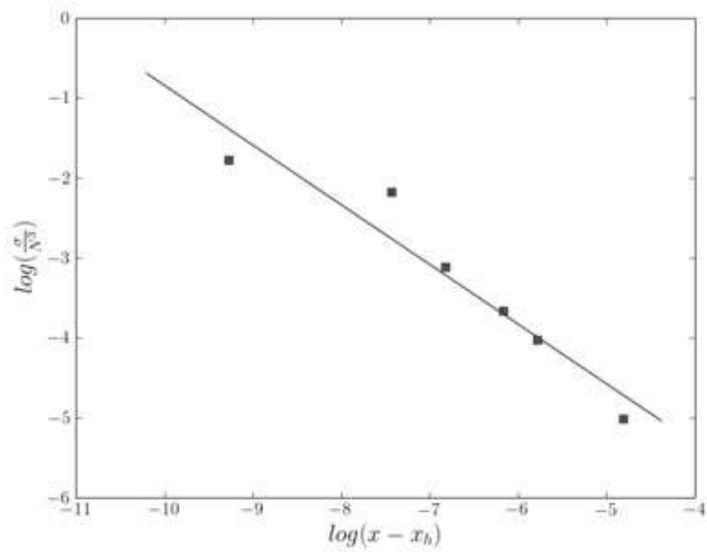

Fig4: Normalized standard deviation ($\sigma/N^3$) in size of the AB clusters as a function of fractional concentration x for sample size N=498.